# Middle-Solving Gröbner bases algorithm for cryptanalysis over finite fields

Wansu Bao[1,2]*, Heliang Huang[1,2]

[1] Zhengzhou Information Science and Technology Institute, Zhengzhou 450000, China

[2] Synergetic Innovation Center of Quantum Information and Quantum Physics, University of Science and Technology of China, Hefei 230026, China

## ABSTRACT

Algebraic cryptanalysis usually requires to recover the secret key by solving polynomial equations. Gröbner bases algorithm is a well-known method to solve this problem. However, a serious drawback exists in the Gröbner bases based algebraic attacks, namely, any information won't be got if we couldn't work out the Gröbner bases of the polynomial equations system. In this paper, firstly, a generalized model of Gröbner basis algorithms is presented, which provides us a platform to analyze and solve common problems of the algorithms. Secondly, we give and prove the degree bound of the polynomials appeared during the computation of Gröbner basis after field polynomials is added. Finally, by detecting the temporary basis during the computation of Gröbner bases and then extracting the univariate polynomials contained unique solution in the temporary basis, a heuristic strategy named Middle-Solving is presented to solve these polynomials at each iteration of the algorithm. Farther, two specific application mode of Middle-Solving strategy for the incremental and non-incremental Gröbner bases algorithms are presented respectively. By using the Middle-Solving strategy, even though we couldn't work out the final Gröbner bases, some information of the variables still leak during the computational process.

## KEYWORDS

Gröbner bases, F4, signature Gröbner bases algorithm, algebraic attacks, field equations, Middle-Solving strategy, multivariate polynomial systems of equations

**\*Correspondence**

Wansu Bao, Zhengzhou Information Science and Technology Institute, Zhengzhou 450000, China

E-mail: glhhl0773@126.com

## 1 INTRODUCTION

As one of the most efficient attacks, algebraic attacks have been successful in breaking several stream ciphers, public key cryptosystems, and a few block ciphers. Algebraic attacks try to reformulate a cipher as a (very large) system of polynomial equations and then find the secret key by solving such a system. In this paper, we focus on the polynomial system solving part. The problem of solving polynomial systems over finite fields is known to be very difficult (non-deterministic polynomial-time hard complete in general). The security of many cryptographic systems is based on this problem, which makes developing algorithms for solving polynomial systems be a hot research topic in cryptanalysis.

Gröbner bases, first introduced in [1], are by now a fundamental tool for tracking this problem and become a powerful method for algebraic attacks. In addition, Gröbner bases can be used to determine optimal equations in terms of degree and/or variables in the algebraic attacks. What's more, Albrecht and Cid [2] use Gröbner bases algorithms to perform a consistency check. This allows them to determine whether given pair satisfies the considered differential characteristic. Cryptanalysis involving the Gröbner bases algorithms has been claimed to attack many cryptosystems: multivariate public key cryptosystems such as HFE [3], Minrank [4], McEliece [5], stream ciphers such as Bivium[6], hash function such as SHA-1 [7].

Finding Gröbner bases is a difficult task, which requires lots of computational resources. Algorithms to compute Gröbner bases have evolved a great deal since the first one was proposed in 1965 by Bruno Buchberger [1]. A significant leap in performance was achieved with the introduction of the F4 [8] and F5 [9] algorithms by Jean-Charles Faugère. In fact, F4 and F5 can be regarded as the two sides of Faugère's algorithm: F4 algorithm uses Gaussian elimination to speed up the time-consuming step of "critical pair" reductions. F5 algorithm uses a powerful criterion to remove useless critical pairs. In recent years, many new variants of F5 are proposed and discussed, for example, F5R[10], Matrix- F5[11], SAGBI-F5[11], F5C[12], F5B[13], F4/5[14], EF5[15], G2V[16], GVW[17], GVWHS[18] and many other algorithms.

In recent years, Gröbner bases algorithms developed rapidly and their computational efficiency has improved significantly. But if we apply them to cryptanalysis, we need to consider the actual needs of cryptanalysis. In cryptanalysis, any information leakages may result in serious threat to cryptosystems. However, a serious drawback exists in the Gröbner bases based algebraic attacks, namely, we won't get any information if we couldn't work out the Gröbner bases of the polynomial equations system. In this paper, firstly, a generalized model of Gröbner basis algorithms is presented, which provides us a platform to analyze and solve common problems of the algorithms. Secondly, we give and prove the degree bound of the polynomials appeared during the computation of Gröbner bases after field polynomials is added, which provides a theoretical basis for the subsequent study. Finally, by detecting the temporary basis during the computation of Gröbner basis and then extracting the univariate polynomials contained unique solution in the temporary basis, a heuristic strategy named Middle-Solving is presented to solve these polynomials at each iteration of the algorithm. Farther, two specific application mode of Middle-Solving strategy for the incremental and non-incremental Gröbner bases algorithms are presented respectively. By using the Middle-Solving strategy, even though we couldn't work out the final Gröbner bases, some information of the variables still leak during the computation process. We must stress that the heuristic strategy of Middle-Solving has never been applied to Gröbner bases algorithms until now. Experiments have been presented to demonstrate that the Middle-Solving strategy has the ability to improve the practical of Gröbner bases algorithms drastically.

The paper is structured as follows. First we do some preliminaries in Sect. 2. In Sect.3 a generalized model of Gröbner basis algorithms is presented. The upper bounds for the degree of the polynomials appear during the computation of Gröbner bases is demonstrated in Sect. 4. In Sect.5 we describe our Middle-Solving strategy and introduce experimental results on various benchmark systems in Sect.6. Sect.7 concludes this paper.

## 2 GRÖBNER BASES AND BUCHBERGER ALGORITHM

This section describes the fundamental notations and the conventions in this paper. We briefly give the main definitions needed to define a Gröbner bases in a characterization useful for our purpose and simply describe the algorithm for computing Gröbner bases.

Let $K$ be a field and $R = K[x_1, x_2, \ldots, x_n]$ be the polynomial ring over the field $K$ with $n$ variables. Let $<_T$ denote a fixed admissible ordering on the monomials of $R$. The leading monomial and leading term of the polynomial $p \in R$ with respect to $<_T$ are denoted by $LM(p)$ and $LT(p)$ respectively, and the set of all monomials in polynomial $p$ is denoted by $T(p)$. A Gröbner bases of $I = \langle F = \{f_1, f_2, \ldots, f_m\} \rangle$ with respect to $<_T$ is a finite list $G$ of polynomials in $I$ that satisfies the properties $\langle G \rangle = I$ and for every $p \in I$ there exists $g \in G$ satisfies $LM(g)|LM(p)$. If, in addition, every $g \in G$ is monic and has no monomial that is divisible by $LM(h)$ for any $h \in G$, then $G$ is a reduced Gröbner bases. Buchberger first found an algorithm to compute such a bases. We describe Buchberger's algorithm in the following way and introduce some definitions at the same time: set $G=F$, then iterate the following two steps.

- Choose a pair $p, q \in G$ that has not yet been considered, and construct its *S-polynomial*

$$Spoly(p,q) = \frac{lcm(LM(p), LM(q))}{LT(p)} \cdot p - \frac{lcm(LM(p), LM(q))}{LT(q)} \cdot q$$

- *Reduce* $Spoly(p,q)$ with respect to $G$. That is, $r_0 = Spoly(p,q)$, and while exist $t \in T(r_i)$ remains divisible by $u = LT(g)$ for some $g \in G$, put $r_{i+1} := r_i - \frac{t}{u} \cdot g$. If the reduction of $Spoly(p,q)$ terminates after $j$ iterations, no more reductions of $r_j$ are possible, denoted $Spoly(f,g) \xrightarrow{G} r_j$. If $r_j = 0$, we say that *Spoly(p,q) reduces to zero with respect to G*. If $r_j \neq 0$, we say that $Spoly(p,q)$ reduces to a normal form $r_j$, and append $r_j$ to $G$.

The algorithm terminates once the *S-polynomials* of all pairs $p, q \in G$ top-reduce to zero.

**Theorem 2.1** [19] Let $F$ be the input of Buchberger algorithm. Then the output $G$ of Buchberger algorithm is a Gröbner bases of $\langle F \rangle$ w.r.t. $<_T$.

In fact, not all of the *S-polynomials* need to be reduced. A *S-polynomial* is call useless if it can be reduced to zero *w.r.t. G*. The computations of these *S-polynomials* are redundant. In 1965, Buchberger introduced the following two criteria to detect useless *S-polynomials* and skip the normal form calculation altogether if the *S-polynomial* meets these two criteria during the computation of Gröbner bases algorithms.

**Theorem 2.2** [1] (Buchberger's First Criterion) Let $f, g \in G \subset K[X]$ be two elements such that $lcm(LM(f), LM(g)) = LM(f) \cdot LM(g)$. Then $Spoly(f,g) \xrightarrow{G} 0$.

**Theorem 2.3** [2] (Buchberger's Second Criterion) Let $f, g, h \in G \subset K[X]$. Assume that

(1) $LM(g)|\ lcm(LM(f), LM(h))$, and

(2) *Spoly*(*f*, *g*) and *Spoly*(*g*, *h*) reduce to zero with respect to G.

Then $Spoly(f,h) \xrightarrow{G} 0$.

In order to avoid more redundant reductions, Faugère introduced the concept of label polynomial and proposed F5 algorithm. Based on the idea of F5 algorithm, a series of algorithms developed. Collectively, we call the Gröbner bases algorithms which act on label polynomial as signature Gröbner bases algorithms. In order to facilitate the distinction, we call the Gröbner bases algorithms which act on normal polynomial as classic Gröbner bases algorithms, such as Buchberger algorithm and F4 algorithm.

# 3 A GENERALIZED MODEL OF THE GRÖBNER BASES ALGORITHM

Following his supervisor's advice, Buchberger used S-polynomials to eliminate the leading term during the computation of Gröbner bases algorithm. The subsequent proposed Gröbner bases algorithms never jump out of the basic idea of Buchberger algorithm essentially. For example, F4 algorithm, signature Gröbner bases algorithm (F5, Matrix-F5, F5B, F5R, F5C, G2V, GVW and so on). This article will not describe these algorithms one by one, instead, by summing up the existing Gröbner bases algorithms, a generalized model of Gröbner bases algorithm is described in algorithm 3.1.

**Algorithm 3.1** A generalized model of Gröbner algorithm

**inputs**: $F=(f_1,\ldots,f_m), G=\varnothing$

**outputs**: The Gröbner bases for $I = \langle F \rangle$

1. **while** $F \neq \varnothing$ **do**
2. $F' := Extract(F)$
3. $G := G \cup F'$
4. $S := S\text{-}polynomial(F',G)$
5. **while** $S \neq \varnothing$ **do**
6. $S' := Select(S)$
7. $\tilde{F} := Reduce(S')$
8. $G := G \cup \tilde{F}$
9. $S := S \cup S\text{-}polynomial(\tilde{F},G)$
10. **return** $poly(G)$

The all existing Gröbner bases algorithms can be described under the framework of algorithm 3.1. The only difference of these algorithms is just that different strategies are used in each sub-algorithm. According to the algorithm 3.1, Gröbner bases algorithm can be divided into input stage, S-polynomial generation stage, reduction stage and output stage. The following we will generally describe each stage of the Gröbner bases algorithm:

**1. Input stage.** Extract the polynomial set $F'$ from initial polynomial $(f_1,\ldots,f_m)$ to execute the following operations, where $f_i \in K[X]$. According to the structure of algorithms, Gröbner bases algorithms can be divided into incremental Gröbner bases algorithm and non-incremental Gröbner bases algorithm. If one wants to compute a Gröbner bases for an ideal $I = \langle f_1,\ldots,f_m \rangle$ in the incremental Gröbner bases algorithms world we compute the Gröbner bases $G_1$ for $\langle f_1 \rangle$, then $G_2$ for $\langle f_1, f_2 \rangle$, and so on until we reach $G_m$, a Gröbner bases for *I*. Most signature Gröbner bases algorithms presented now depend on incremental computations. For example, F5、G2V、F5C and so on. Unlike incremental algorithms, non-incremental Gröbner bases algorithms compute the Gröbner bases of $\langle f_1,\ldots,f_m \rangle$ directly. Some of signature Gröbner bases algorithms (such as EF5, F5B, GVW, etc.), as well as classic Gröbner bases algorithms (such as F4, Buchberger algorithm, etc.) have adopted a non-incremental structure. Line 2 is used to distinguish the incremental and non-incremental structure. If the algorithm is incremental structure, then *Extract* represents to extract a $f_i, i \in 1,\ldots,m$ from $F$. If the algorithm is non-incremental structure,then *Extract* represents to extract $F$.

**2. S-polynomial generation stage.** Line 4 and 9 are used to generate S-polynomials. In this stage, some criteria can be used to detect useless critical pairs during the computation of Gröbner bases. If we don't apply criteria to avoid generating redundance S-polynomials $S\text{-}polynomial(F,G) := \{Spoly(f_i, f_j) \mid f_i, f_j \in F, f_i \neq f_j\}$ $\bigcup \{Spoly(f_i, g_j) \mid f_i \in F, g_j \in G, f_i \neq g_j\}$. If some criteria are used, the S-polynomials satisfing the criteria in the set of $S\text{-}polynomial(F,G)$ could be deleted.

**3. Reduction stage.** Line 6 and 7 are used to select and then reduce S-polynomials. Now, the strategy that reducing critical pairs/S-polynomials with the smallest degree first is commonly used. Faugère has said, during the computation of a Gröbner bases, almost all time are spent on reducing polynomials. Thus, speeding up the efficiency of reduction stage will improve efficiency of the whole algorithm significantly. Combining with the matrix technique is a very effective way to speed up the efficiency of reduction stage.

**4. Output stage.** Line 1, 5 and 10 can be regarded as the output stage. Line 1 and 10 are used to determine whether $G$ is the Gröbner bases of $I = \langle f_1,\ldots,f_m \rangle$. When $F=\emptyset$ and $S=\emptyset$ are both satisfied, then output $poly(G)$, the Gröbner bases of $I = \langle f_1,\ldots,f_m \rangle$. All Gröbner bases algorithms are iterative algorithms, we call $G$ appears in Line 1-9 as a temporary bases. In the incremental Gröbner bases algorithms, when $F \neq \emptyset$ and $S=\emptyset$, temporary bases $G$ is the Gröbner bases of $\langle f_1,\ldots,f_i \rangle$ (or $\langle f_i,\ldots,f_m \rangle$). At this time, we denote the algorithm has completed a round of iteration and then extract the next $f_{i+1}$ (or $f_{i-1}$) to compute the Gröbner bases of $\langle f_1,\ldots,f_{i+1} \rangle$ (or $\langle f_{i-1},\ldots,f_m \rangle$). Until $F=\emptyset$ and $S=\emptyset$, the algorithm terminates. In the non-incremental Gröbner bases algorithms, we denote the algorithm has completed a round of iteration when the algorithm has executed Line 9 a time.

Figure 3.1 shows the flow chart of Gröbner bases algorithm, the paper will do some research for the input stage and output stage.

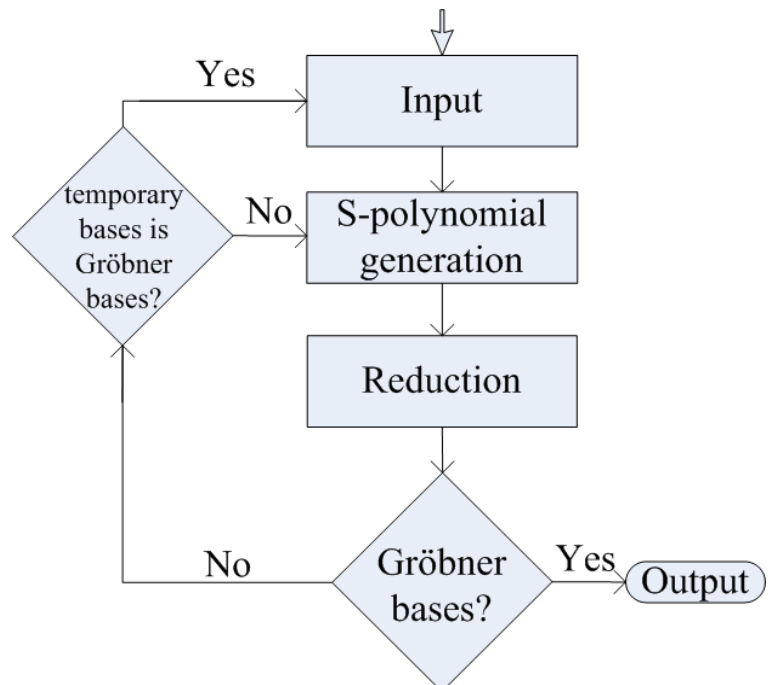

Figure 3.1. The flow chart of Gröbner bases algorithm

# 4 THE UPPER BOUNDS FOR THE DEGREE OF POLYNOMIALS APPEAR IN GRÖBNER BASES ALGORITHM

For cryptographic purpose, solutions in the algebraic closure are irrelevant for us. Usually, only solutions over the finite fields are of importance. In Gröbner bases algorithm, a potential way to deal with this issue is to try to adjoin the set of field equations to the list of equations that we want to solve. For a polynomial ring $R=F_q[X]$ we can write the set of field equations of the form $x^q+x=0$ for all $x \in X$, where $X=\{x_1,\ldots,x_n\}$. It is equivalent to computing Gröbner bases of $\langle f_1,\ldots,f_m,x_1^q+x_1,\ldots,x_n^q+x_n \rangle$. By adjoining the set of all field polynomials to the initial set of polynomials $I$, all variables in the final Gröbner bases can be force to satisfy the field equations $\{x^q+x \mid x \in X\}$.

In fact, the strategy which taking the field equations into account for Gröbner bases algorithm has also been used previously. We mention it here, it is just because we want to discuss the regular for the degree of polynomials, monomials and variables appear during the computation of Gröbner bases algorithm after adding the field equations.

**Lemma 4.1** Let polynomial ring $R=F_q[X]$, $X=\{x_1,\ldots,x_n\}$. By adjoining the set of field polynomials $F=\{x^q+x \mid x \in X\}$ to the initial polynomial ideal, the maximal degree of all polynomials in the temporary bases and final output Gröbner bases during the computation of Gröbner bases algorithm is at most $n(q-1)$, and the degree of every variable in the leading term is at most $q$.

*proof.* The temporary bases or final output Gröbner bases contains the field polynomials and S-polynomials reduced completely. The field polynomials are consistent with lemma 4.1. Assume, for contradiction, that exists polynomial $f$, a S-polynomial reduced completely, in the temporary bases or final output Gröbner bases such that $\deg(f) > n(q-1)$. Without loss of generality, assume there are $m$ variables in $LT(f)$, that is $x_i \in \{x_{i_1}, x_{i_2}, \ldots, x_{i_m}\}$, where $m \le n$. Denote the degree of variable $x_i$ in $LT(f)$ as $power(x_i)$, where $x_i \in \{x_{i_1}, x_{i_2}, \ldots, x_{i_m}\}$. Due to $\deg(LT(f)) = \deg(f) > n(q-1)$, so there must exist a variable $x_i \in \{x_{i_1}, x_{i_2}, \ldots, x_{i_m}\}$ such that

$$power(x_i) > \frac{n(q-1)}{m} \geq \frac{n(q-1)}{n} = q-1$$

If there is a variable $x_i$ in $f$ such that $power(x_i) > q-1$, we could continue to use $x_i^q + x_i$ to reduce $f$, This is a contradiction to the hypothesis that $f$ is reduced completely.

According to the proof of Lemma 4.1, it is easy to know that the degree of every variables in the leading term of S-polynomials reduced completely in temporary bases or final output Gröbner bases is lower than $q$, and the degree of every variables in the leading term of field polynomials is $q$. The polynomials appear during the computation of Gröbner bases algorithm can be divided into S-polynomial, reductor and the polynomials in the temporary bases or final output Gröbner bases. Next, we will discuss the upper degree bound of these polynomials.

**Theorem 4.1** Let polynomial ring $R = F_q[X]$, $X = \{x_1, \ldots, x_n\}$. By adjoining the set of field polynomials $F = \{x^q + x \mid x \in X\}$ to the initial polynomial ideal, the maximal degree of all polynomials appear during the computation of Gröbner bases algorithm is at most $n(q-1)+1$.

*proof.* According to the Lemma 4.1, the maximal degree of all polynomials in the temporary bases and final output Gröbner bases is at most $n(q-1)$. Next we will prove the highest degree of S-polynomials and reductors is $n(q-1)+1$. The computational formula of S-polynomial is

$$Spoly(f,g) := \frac{lcm(LM(f), LM(g))}{LT(f)} f - \frac{lcm(LM(f), LM(g))}{LT(g)} g,$$

where $f$ and $g$ are taken from the temporary bases.

**Case a.** If both $f$ and $g$ are S-polynomials reduced completely in the last round, then the degree of every variables in $LM(f)$ and $LM(g)$ is lower than $q$. It is easy to know that the degree of every variables in $lcm(LM(f), LM(g))$ is lower than $q$, So the highest degree of $lcm(LM(f), LM(g))$ is $n(q-1)$. According to the computational formula of S-polynomial, we can obtain that the degree of the S-polynomials is at most $n(q-1)$.

**Case b.** If $f$ is a S-polynomial reduced completely in the last round and $g$ is a field polynomial, then the degree of every variables in $LM(f)$ is lower than $q$ and $LM(g) = x_i^q$, where $x_i \in X$. It is easy to know that the degree of every variables in

$lcm(LM(f), LM(g))$ is lower than $q$ except the degree of $x_i$ is equal $q$. So the degree of the S-polynomials is at most $n(q-1)+1$.

Above all, the degree of the S-polynomials is at most $n(q-1)+1$. Since the leading term of reductor is equal to a certain monomial in S-polynomial, the degree of the reductors is at most $n(q-1)+1$.

Of particular interest in the case $q=2$, where we can easily have the following simple corollary.

**Corollary 4.1** Let polynomial ring $R = F_2[X]$, $X = \{x_1, \ldots, x_n\}$. By adjoining the set of field polynomials $F = \{x^2 + x \mid x \in X\}$ to the initial polynomial ideal, the maximal degree of all polynomials appear during the computation of Gröbner bases algorithm is at most $n+1$.

## 5 MIDDLE-SOLVING GRÖBNERBASES ALGORITHM

In this section, we will slightly modify the logic of the Gröbner bases algorithms to make them be more practical for cryptanalysis (especially algebraic attacks) to solve cryptosystems over finite fields.

Let initial polynomial ideal $I = \langle F \rangle$, $F \subset F_q[X]$. A Gröbner bases for a lexicographical order (Lex) of a zero-dimensional system (i.e. with a finite number of solutions) has the following shape

$$\{g_1(x_1), \ldots, g_2(x_1, x_2), \ldots, g_{k_1}(x_1, x_2), g_{k_1+1}(x_1, x_2, x_3), \ldots, g_{k_n}(x_1, \ldots, x_n)\} \cdots\cdots\cdots ①$$

**Theorem 5.1** [20] If the solutions of $\{f_1(x_1, \ldots, x_n) = 0, \ldots, f_m(x_1, \ldots, x_n) = 0\}$ is limited, a Gröbner bases for a lexicographical order of $I = \langle f_1, \ldots, f_m \rangle$ has a triangular structure.

With such structure, solutions can be easily computed by successively eliminating variables, namely computing solutions of univariate polynomials and back-substituting the results. The basic idea of using Gröbner bases algorithms to solve equations can be illustrated as in Figure 4.1: In order to solve equations $\{f = 0 \mid f \in F\}$, firstly, we should compute the Gröbner bases of the initial polynomials system $I = \langle F \rangle$, then we get the value of the variables from the Gröbner bases by other algorithm (e.g. with Berlekamp's algorithm). The solution of Gröbner bases is the solution of the equations $\{f = 0 \mid f \in F\}$.

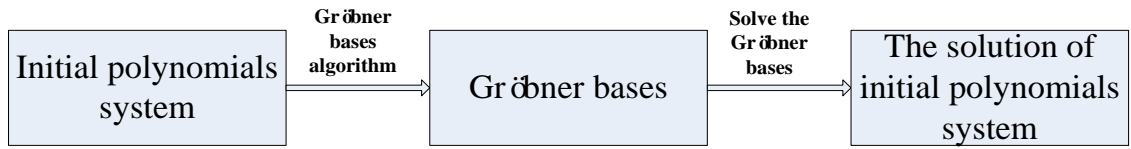

Figure 4.1 The steps of Gröbner bases algorithm for solving nonlinear equations

The ultimate aim of algebraic attacks is to obtain the solution of the initial equations system $\{f = 0 \mid f \in F\}$, but not the Gröbner bases of the initial equations system. Gröbner bases just can be regard as an intermediate step to solve systems of polynomial equations symbolically. In cryptanalysis, any information leakages may result in serious threat to cryptosystems. However, it is not easy to work out the Gröbner bases of a large-scale cryptographic equations system. So, a serious drawback exists in the Gröbner bases based algebraic attacks, namely, we won't get any information if we couldn't work out the Gröbner bases of the polynomial equations system. This drawback greatly restricts the practicability of algebraic attacks.

By adjoining the set of field polynomials to the initial polynomial ideal, Gröbner bases algorithms can be controlled to run at a low degree. In particular, the upper bound for the degree of polynomials in the temporary bases is just $n+1$ after adjoining the set of field polynomials $F = \{x^2 + x \mid x \in X\}$ over GF(2). By detecting the polynomials in the temporary bases during the computation of Gröbner bases algorithm, we have observed there are some univariate polynomials that had only one solution appear in the temporary bases. They are just treated like other polynomials in the Gröbner bases algorithms. Due to the limitation of time or memory, if we couldn't work out the final Gröbner bases, unfortunately, then these polynomials can't be found. At this time we will get nothing about the solutions. However, these polynomials can provide part of the value of the solutions, so they should be deserved a special treatment during the computation of Gröbner bases algorithms.

To solve this problem, we add a detection algorithm into Gröbner bases algorithm to search those univariate polynomials that had only one solution in the temporary bases during the computation of Gröbner bases algorithm. If exists, solve these univariate polynomials and back-substitute the values of the solved variables. Once a value of variable is obtained, the following operation is equivalent to computing Gröbner bases with respect to the unsolved variables. We name the heuristic strategy Middle-Solving strategy. We mention that our heuristic strategy, by design, will boost the practicability of all Gröbner bases algorithms for solving equations. Even though we couldn't work out the final Gröbner bases, some information of the variables still leak during the computation process. In addition, after back-substituting the values of the solved variables, the algorithm is equivalent to solving Gröbner bases *w.r.t* remaining variables and polynomials, which makes the whole algorithm get relatively simpler than before.

**Theorem 5.2** [21] Let $I = \langle f_1, \ldots, f_m \rangle$, $G$ be the reduced Gröbner bases of ideal $I$. If exists $1 \in G$, then equations $\{f_1(x_1, \ldots, x_n) = 0, \ldots, f_m(x_1, \ldots, x_n) = 0\}$ has no solution.

**Theorem 5.3** Let polynomial ring $R = F_q[X]$, $X = \{x_1, \ldots, x_n\}$. Assume the imput of Gröbner bases algorithm is $(f_1, \ldots, f_m)$. Let $I = \langle f_1, \ldots, f_m \rangle$, $G'$ is the temporary bases during the computation of Gröbner bases algorithm, $G$ is the reduced Gröbner bases of $I$. If exists $k \in G'$, where $k \neq 0 \in F_q$, then $1 \in G$.

*proof.* If $k \in G'$, then $k$ can be interreduced to 1 [22]. That is to say $1 \in G$.

With Theorem 5.2 and 5.3, Corollary 5.1 is proved.

**Corollary 5.1** Let polynomial ring $R = F_q[X]$, $X = \{x_1, \ldots, x_n\}$. Assume the imput of Gröbner bases algorithm is $(f_1, \ldots, f_m)$. Let $I = \langle f_1, \ldots, f_m \rangle$, $G'$ is the temporary bases during the computation of Gröbner bases algorithm, $G$ is the reduced Gröbner bases of $I$. If exists $k \in G'$, where $k \neq 0 \in F_q$, then equations $\{f_1(x_1, \ldots, x_n) = 0, \ldots, f_m(x_1, \ldots, x_n) = 0\}$ has no solution.

In particular, $k$ equals to 1 only over GF(2). So as long as $1 \in G'$, then equations $\{f_1(x_1, \ldots, x_n) = 0, \ldots, f_m(x_1, \ldots, x_n) = 0\}$ has no solution. Based on Corollary 5.1 we could detect whether there is $k \neq 0 \in F_q$ in temporary bases. If exists, it indicates that the input equations has no solution. That is to say we no longer need to calculate the Gröbner bases of input. Then the algorithm can be directly terminated. This will save a lot of useless calculations. So we also include this detection algorithm in the Middle-Solving strategy.

In short, Middle-Solving strategy can allow attackers to obtain the information of solution as more as possible when using Gröbner bases algorithm to solving equations. According to Figure 3.1, a flowchart of the Middle-Solving Gröbner bases algorithm is presented in Figure 5.1.

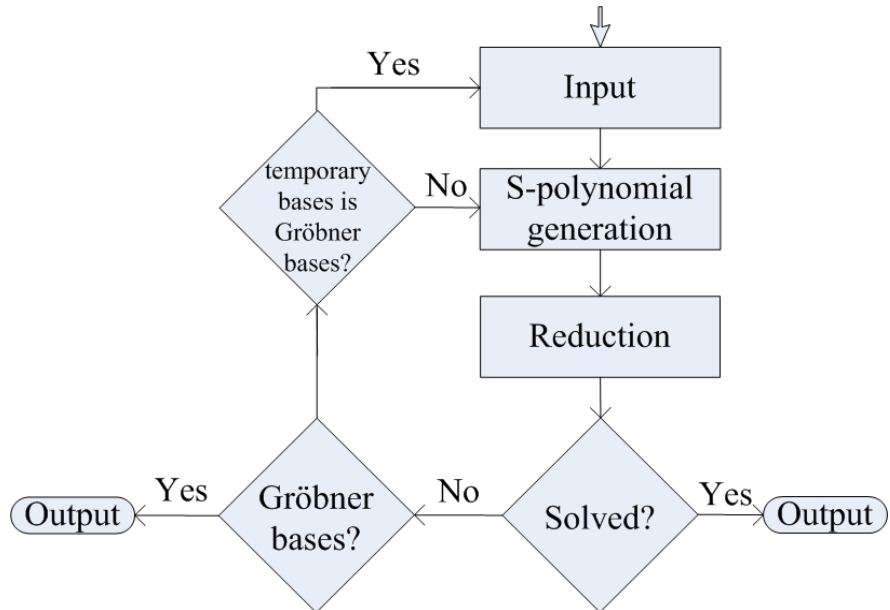

Figure 5.1. The flowchart of Middle-Solving Gröbner bases algorithm

# 6 EXPERIMENTAL RESULTS AND ANALYSIS

In this section, two specific application modes of Middle-Solving strategy for the incremental and non-incremental Gröbner bases algorithms are presented respectively. Experimental results are presented to compare Middle-Solving Gröbner bases algorithm to original Gröbner bases algorithm for these two application modes. We take an interest in solving systems of some classical benchmarks (Cyclic6, Gonnet83 and so on). The implementations in this section are all written by Magma version (V2.11-11).

## 6.1 The Application of Middle-Solving Strategy in Non-incremental Gröbner Bases Algorithms

Classic Gröbner bases algorithms (F4, Buchberger algorithm) and some signature Gröbner bases algorithms (EF5, F5B, GVW and so on) are based on a non-incremental frame. F4 algorithm is recognized as one of the most efficient algorithms. Here we take F4 algorithm as an example to illustrate how to apply Middle-Solving strategy into non-incremental Gröbner bases algorithms. Obviously we can use it to boost the performance of all members of non-incremental Gröbner bases algorithms in the same way as it aids F4 algorithm.

Non-incremental Gröbner bases algorithms computer the Gröbner bases of $I = \langle f_1, \ldots, f_m \rangle$ one shot. Middle-Solving strategy can be applied to the set of reduced

S-polynomials after each iteration. The main loop of Middle-Solving F4 is presented in Algorithm 6.1, where line 10-17 and line 20-21 are the pseudo-code description of Middle-Solving strategy. Other sub-algorithms are consistent with the Faugère's description. (Readers can refer to [8] for the complete pseudo code of F4 algorithm).

**Algorithm 6.1** Middle-Solving F4

**inputs:** $F = (f_1, f_2, \ldots, f_m) \in R^m$
**Initialization:** $G := \emptyset$, $P := \emptyset$, $\mathbb{F} := 0$, $d := 0$
1:  **while** $F \neq \emptyset$ **do**
2:      $f := first(F)$
3:      $F := F \setminus \{f\}$
4:      $(G, P) := Update(G, P, f)$
5:  **while** $P \neq \emptyset$ **do**
6:      $d := d+1$
7:      $P_d := Select(P)$
8:      $P := P \setminus P_d$
9:      $(\tilde{F}_d^+, F_d) := Reduction(P_d, G, (F_i)_{d=1,\ldots,(d-1)})$
10:     $UP := [f : f \text{ in } \tilde{F}_d^+ \mid IsUnivariate(f)]$
11:     **if** $UP$ is not Empty **then**
12:         $R := [f : f \text{ in } UP \mid \#Roots(UnivariatePolynomial(f)) \text{ eq } 1]$
13:         **if** $R$ is not Empty **then**
14:             **for** $r$ in $R$ **do**
15:                 $root := Solve(UnivariatePolynomial(r))$
16:                 $PrintFile(root)$       //Output the the values of the solved variables
17:                 $Renew(G, \tilde{F}_d^+, \mathbb{F})$       //Back-substitute the values of the solved variables
18:     **for** $h \in \tilde{F}_d^+$ **do**
19:         $(G, P) := Update(G, P, h)$
20:     **if** $\exists k \neq 0 \in F_q, s.t. k \in G$ **then**
21:         **break**
22: **return** $G$

Line 10-13 are used to detect whether there is univariate polynomials that had only one solution in $\tilde{F}_d^+$, the set of reduced S-polynomials. If exists, Line 14-17 are used to solve these univariate polynomials, and then output and back-substitute the values of the solved variables. In Line 20-21 if exists $k \in G$, where $k \neq 0 \in F_q$, it indicates that the input equations $\{f_1 = 0, f_2 = 0, \ldots, f_m = 0\}$ has no solution. Then we can break the algorithm. In particular, $k$ equals to 1 only over GF(2).

Experimental results to compare Middle-Solving F4 with the original F4 for some classical benchmarks over GF(2) are presented in Table 6.1. "$n$" denotes the number of variables in the input equations. "Round" and "#Solved" in the tuple "(Round,#Solve)" represent the iteration round of the algorithm and the number of solved variables when Middle-Solving strategy detects univariate polynomials that had only one solution in $\tilde{F}_d^+$. The total iteration round of the algorithm is represented by "Total Round". Experiments show that Middle-Solving strategy can effectively detect the univariate polynomials that had only one solution during the computation of Gröbner bases algorithm and then output the values of the solved variables. Even though we couldn't work out the final Gröbner bases, some information of the variables still leak during the computation process. Meanwhile, using Middle-Solving strategy may get all the values of variables during the computation of the algorithm, which makes the algorithm terminates with fewer rounds.

For example, for Eco12, Midlle-Solving F4 can detect and then solve two univariate polynomials at 11th round, and gets all the values of variables and then terminates at 13th round. Original F4 terminates at 14th round. Assume the algorithm can only run 12 rounds due to storage overflow, unfortunately original F4 will get nothing about the solution. However, Midlle-Solving F4 still can obtain the value of the two variables.

Table 6.1. Performance of Middle-Solving F4 versus original F4 for some benchmarks over GF(2)

| Test | $n$ | (Round,# Solve) Middle-Solving F4 | Total Round F4 | Middle-Solving F4 |
|---|---|---|---|---|
| Cyclic8 | 8 | (11,8) | 14 | 11 |
| Katsura-10-h | 11 | (2,4) | 4 | 4 |
| Eco12 | 12 | (11,2),(13,12) | 14 | 13 |
| Gonnet83 | 7 | (1,7) | 6 | 1 |
| SchransTroost | 8 | (1,8) | 1 | 1 |

## 6.2 The Application of Middle-Solving Strategy in Incremental Gröbner Bases Algorithms

Currently, most signature Gröbner bases algorithms (F5, F5C, G2V and so on) are based on incremental frame. If one wants to compute a Gröbner basis for an ideal $\langle f_1,...,f_m \rangle$ in the incremental Gröbner bases algorithms world we compute the Gröbner bases $G_1$ for $\langle f_1 \rangle$, then $G_2$ for $\langle f_1, f_2 \rangle$, and so on until we reach $G_m$, a Gröbner basis for $\langle f_1,...,f_m \rangle$. So we use Middle-Solving strategy to detect $G_i$, a Gröbner basis for $\langle f_1,...,f_i \rangle$, where $i \in [2,...,m]$. A generalized model of Middle-Solving incremental Gröbner algorithm is presented in Algorithm 6.2, where line 6-15 are the pseudo-code description of Middle-Solving strategy.

**Algorithm 6.2** A generalized model of Middle-Solving incremental Gröbner algorithm

**inputs:** $F = (f_1, f_2, \ldots, f_m) \in R^m$

1: $G_1 := \{f_1\}$
2: **for** $i=2,\ldots,r$ **do**
3:    $f_i := Reduce(f_i, G_{i-1})$
4:    **if** $f_i \neq 0$ **then**
5:       $G_i := IncSiG(f_i, G_{i-1})$       // Compute the reduced Gröbner bases of $(f_i, G_{i-1})$
6:       $UP := [f : f \text{ in } G_i \mid \text{IsUnivariate}(f)]$
7:       **if** $UP$ is not Empty **then**
8:          $R := [f : f \text{ in } UP \mid \#Roots(\text{UnivariatePolynomial}(f)) \text{ eq } 1]$
9:          **if** $R$ is not Empty **then**
10:           **for** $r$ in $R$ **do**
11:              $root := Solve(\text{UnivariatePolynomial}(r))$
12:              $PrintFile(root)$     // Output the values of the solved variables
13:              $Renew(G_i, Rule)$     //Back-substitute the values of the solved variables
14:       **if** $\exists k \neq 0 \in F_q, s.t. k \in G$ **then**
15:          **break**
16:    **else**
17:       $G_i := G_{i-1}$
18: **return** $G_i$

Experimental results to compare Middle-Solving incremental Gröbner bases algorithm(short for "M-S") with the original incremental Gröbner bases algorithm(short for "Increase") for some classical benchmarks over GF(2) are presented in Table 6.2. Experiments show that Middle-Solving strategy is also applicable to incremental Gröbner bases algorithm, which can effectively discover the univariate polynomials during the computation of Gröbner bases algorithm and then output the values of the solved variables.

Table 6.2. Performance of Middle-Solving incremental Gröbner bases algorithm versus original incremental Gröbner bases algorithm for some benchmarks over GF(2)

| Test | $n$ | (Round,#GB) M-S | Total Round Increase | M-S |
|---|---|---|---|---|
| Cyclic8 | 8 | (9,8) | 16 | 9 |
| Katsura-10-h | 11 | (15,1),(17,2),(19,3),(21,4) | 21 | 21 |
| Eco12 | 12 | (24,12) | 24 | 24 |
| Gonnet83 | 7 | (27,1),(28,2),(29,3),(34,7) | 36 | 36 |
| SchransTroost | 8 | (9,1),(10,2),(11,3),(12,4),(13,5),(14,6),(15,7),(16,8) | 16 | 16 |

## 7 CONCLUSION AND FUTRUE WORK

In order to overcome the serious drawback of the Gröbner bases based algebraic attacks that no information leak if we couldn't work out the Gröbner bases of the polynomial equations system, a heuristic strategy named Middle-Solving strategy is presented in this paper. Experimentally, Middle-Solving strategy can effectively discover the univariate polynomials during the computation of Gröbner bases algorithm and then output the values of the solved variables. It indicates that even though we couldn't work out the final Gröbner bases, some information of the variable still leak during the computation process. So this heuristic strategy is well adapted for algebraic attacks on cryptosystems.

**Author contributions:** Wansu Bao drafted the preliminary proposal, the theoretical derivation and programming were carried out by Heliang Huang. They contributed equally to this work.

**Acknowledgements:** This work was supported by the National Basic Research Program of China under Grant No 2013CB338002.